\documentclass[a4paper]{JHEP3}
 
\usepackage{amsmath,amsfonts,a4,epsfig,graphicx}
\numberwithin{equation}{section}
\def\beq{\begin{equation}}
\def\eeq{\end{equation}}
\def\bea{\begin{eqnarray}}
\def\eea{\end{eqnarray}}

\def\lsim{\mathrel{\raisebox{-.6ex}{$\stackrel{\textstyle<}{\sim}$}}}
\def\gsim{\mathrel{\raisebox{-.6ex}{$\stackrel{\textstyle>}{\sim}$}}}

\preprint{KA-TP-01-2008, KEK-TH-1217, KIAS-P08006, SFB/CPP-08-01, UFIFT-HEP-08-01}

\title{ Graviton production with 2 jets at the LHC in large extra dimensions }

\author{ Kaoru Hagiwara$^{a}$, Partha Konar $^{b,c,1}$, Qiang Li$^{c,2}$, Kentarou Mawatari $^{d,3}$, and Dieter Zeppenfeld$^{c,4}$ \\
\\$^a$ KEK Theory Division and Sokendai, Tsukuba 305-0801, Japan \\
$^b$ Institute for Fundamental Theory, University of Florida, Gainesville, FL 32611, USA\\
$^c$ Institut f\"ur Theoretische Physik, Universit\"at  Karlsruhe, Postfach 6980, D-76128 Karlsruhe, Germany\\
$^d$ School of Physics, Korea institute for Advanced Study, Seoul 130-722, Korea\\
$~$ \\ $^1$ E-mail:  konar@phys.ufl.edu  \\
$^2$ E-mail:qliphy@particle.uni-karlsruhe.de \\
$^3$ E-mail:kentarou@kias.re.kr \\
$^4$ E-mail:dieter@particle.uni-karlsruhe.de }

\abstract
{We study Kaluza-Klein (KK) graviton production in the large extra
dimensions model via 2 jets plus missing transverse momentum
signatures at the LHC. We make predictions for both the signal and
the dominant $Zjj$ and $Wjj$ backgrounds, where we introduce missing
$P_T$-dependent jet selection cuts that ensure the smallness of the
2-jet rate over the 1-jet rate. With the same jet selection cuts,
the distributions of the two
jets and their correlation with the missing transverse momentum
provide additional evidence for the production of an invisible
massive object.}

\date{\Date}


\maketitle
\newpage
\begin{document}
%%%%%%%%%%%%%% Begin Main Part %%%%%%%%%%%%%%%%%%%%%%%%%%%%%%%%%%%%%%%%%
%%%%%%%%%%%%%% Begin Section 1 %%%%%%%%%%%%%%%%%%%%%%%%%%%%%%%%%%%%%%%%%

\section{Introduction}\label{sec:1}
A general expectation in high energy physics today is that physics
beyond the standard model (BSM) should emerge at TeV energies. This
belief is founded on the observation that the electroweak symmetry
breaking scale of the SM cannot be made stable against quantum
corrections without invoking new physics at the TeV scale. With this
in mind, an enormous international effort is being poured into the
construction of the 14-TeV Large Hadron Collider (LHC) at CERN.
Apart from supersymmetry (SUSY), models with large extra space
dimensions, such as the one proposed by Arkani-Hamed, Dimopoulos, 
and Dvali (ADD)~\cite{ADD}, provide an alternative
possibility in this direction that is most exciting.

In the $D=4+\delta$ dimensional ADD model, the SM particles live in
the usual $3+1-$dimensional space, while gravity can propagate into
the additional $\delta$-dimensional space, which is assumed for
simplicity to be compactified on the $\delta$-dimensional torus
$T^\delta$ with a common radius $R$. Then the 4-dimensional Planck
scale $M_{Pl}$ is related to the fundamental scale $M_s$ as
follows~\cite{ADD}:
\begin{eqnarray}\label{scale}
M^2_{Pl}=8\pi R^\delta M^{\delta+2}_s ,
\end{eqnarray}
where $M_s\sim {\rm TeV}$ is possible for large compactification
radius $R$. According to Eq.~(\ref{scale}), one can expect that
deviations from the usual Newtonian gravitational force law will
appear at a distance around $R\sim$ $0.83 \times
10^{-16+\frac{30}{\delta}}{\rm mm}(2.4\,{\rm
TeV}/ M_s)^{1+\frac{2}{\delta}}$. Terrestrial experiments
gave the limit $R\leq 0.2$~mm by probing gravitational forces 
directly~\cite{grav}. For $\delta=2$ this translates into $M_s \gtrsim 1.5$~TeV, while 
for $\delta>2$, there are no strong limits on $M_s$.

In our four dimensional space-time, there appear Kaluza-Klein (KK)
towers of massive spin-2 gravitons in the ADD model which interact
with the SM fields. The interaction Lagrangian is given
by~\cite{Feynr1,Feynr2}
\begin{eqnarray}\label{IL}
 {\cal L}_{int} = - \frac{1}{{\overline M}_{Pl}}
\sum_{\vec{n}} G^{(\vec{n})}_{\mu \nu} T^{\mu
\nu},
\end{eqnarray}
where the massive gravitons are labeled by a $\delta$-dimensional
vector of positive integers,
$\vec{n}=(n_1,n_2,..,n_\delta)$, ${\overline M}_{Pl}=M_{Pl}/\sqrt{8
\pi}\sim 2.4\times 10^{18}$~GeV is the reduced four dimensional
Planck scale, and $T_{\mu \nu}$ is the energy-momentum tensor of the
scattering fields. The $\vec{n}$-th KK mode graviton mass squared is
$m_{(\vec{n})}^2= |\vec{n}|^2/R^2$. From Eq.~(\ref{IL}) one can
derive the relevant Feynman rules, some of which can be found in
Ref.~\cite{Feynr1,Feynr2}. For $M_s=1$~TeV and $\delta=$4, 6 and 8,
the mass gap of the KK modes is $\Delta m=R^{-1} \simeq$ 20~keV,
7~MeV and 0.1~GeV~\cite{Feynr1}, respectively. Thus the spectrum of KK
modes can be treated as continuous for $\delta\leq 6$, and
approximated by the mass density function~\cite{Feynr1}
\begin{eqnarray}
\rho(m)= S_{\delta-1}\frac{{\overline
M}_{Pl}^2}{M_s^{2+\delta}}m^{\delta-1},\, \,\, \, \rm{with}\, \,
S_{\delta-1}=\frac{2\pi^{\delta/2}}{\Gamma(\delta/2)}.
\end{eqnarray}

So far the strongest constraints for $\delta<4$ extra dimensions
come from astrophysics and cosmology, but they can be relaxed and do
not diminish the importance of collider phenomenology~\cite{PDG}. We
therefore discuss the $\delta=3$ case as well. In collider
experiments, there are two classes of effects that can probe large
extra dimensions: virtual KK tower exchange between the SM particles
and a real graviton emission. Since the couplings of the graviton
with matter are suppressed  by inverse power of ${\overline
M}_{Pl}$, graviton direct production leads to missing energy
signals. Detailed studies have been performed at LEP~\cite{LEP} and
the Tevatron~\cite{TEVATRON}, by searching for gravitons in the
processes $e^+e^-\rightarrow \gamma (Z) +E^{{\rm miss}}$ and 
$p\bar{p}\rightarrow \gamma ({\rm jet}) +P_T^{\rm miss}$.
The combined LEP 95$\%$ CL limits are $M_s>$ 1.60, 1.20, 0.94, 0.77, 0.66~TeV 
for $\delta=$ 2,$\cdots$,6 respectively,
and the Tevatron 95$\%$ CL limits are $M_s>$ 1.18, 0.99, 0.91, 0.86, 0.83~TeV.
For the LHC, graviton production with a monojet has been 
investigated in detail and found to have strong ability to probe up 
to much higher extra dimension scale~\cite{LHCG}.
There is very little information on the underlying physics, however, 
in the transverse momentum and the rapidity of the single jet. One may thus 
wonder whether additional jets in graviton production can be used as a 
more sophisticated probe.

In this paper, we study graviton production with 2 jets at the LHC
in large extra dimensions, in comparison with the dominant $Zjj$
background~\cite{vjj}, and examine if the 2-jet rate and
correlations can give us more information about the mass scale of
the missing object, in addition to the missing $P_T$ distribution.
At the same time we address the questions to what extent high 
transverse momentum graviton production will indeed emerge as a 
monojet signature. We will show that higher order QCD effects do 
lead to a more complex event structure, with multiple jets in the 
100~GeV $P_T$ range.
The remainder of this work is organized as follows: In Section~II we
present the calculations. In Section~III we give numerical results
and discussions. Section~IV contains our conclusions.

%%%%%%%%%%%%%% Begin Section 2 %%%%%%%%%%%%%%%%%%%%%%%%%%%%%%%%%%%%%%%%%

\section{Calculations}
\label{sec:2} We are considering the QCD production of a graviton
with 2 jets at the LHC, $pp\rightarrow jjG_n$, including all the
possible subprocesses, among which $gg\rightarrow ggG_n$,
$gq\rightarrow gqG_n$ and $qq^{(')}\rightarrow qq^{(')}G_n$ play the
most important role. Representative Feynman diagrams are shown in
Fig.~\ref{F1}. In addition to the QCD processes of Fig.~1, we have
also calculated the electroweak (EW) contributions to $jjG_n$
production. In particular, we have determined the graviton production
cross sections from weak boson fusion (WBF) processes. However, the
WBF cross sections for $jjG_n$ production represents a small
correction, which is below 1$\%$, even when imposing typical cuts to
enhance WBF over QCD sources~\cite{Eboli:2000ze}. Thus, WBF processes
do not appear as a promising avenue for studying graviton production
at the LHC, and we do not include them in the results below.

Significant background can come from any
processes leading to  two jets and missing transverse momentum,
among which we consider the most important one, namely $Zjj$
production with subsequent decay 
$Z\rightarrow \nu\bar{\nu}$~\cite{vjj}. We also studied another 
class of processes which could be
significant at least when missing $P_T$ is not too large. This can
arise from QCD production of $Wjj$ with subsequent decay $W^\pm
\rightarrow l^\pm{\nu}$ when the charged leptons $l=e,\mu,\tau$ are
not identified. Here we follow the procedure of Ref.~\cite{Eboli:2000ze}.

\FIGURE{
\centering \epsfig{file=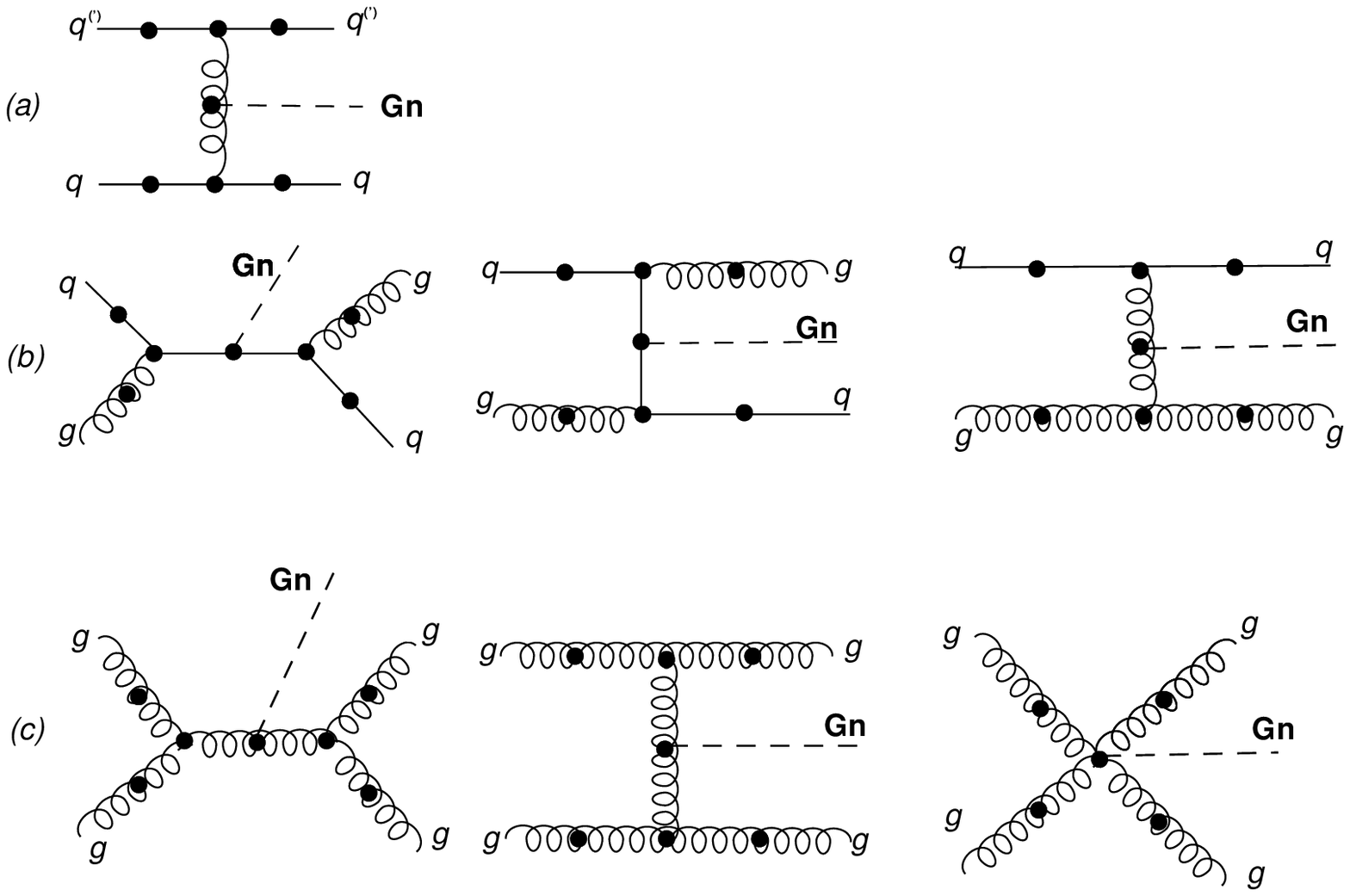,width=0.6\columnwidth}
\caption{\label{F1} Representative Feynman diagrams for the
subprocesses (a) $qq^{(\prime)}\rightarrow qq^{(\prime)} G_n$,
(b) $qg\rightarrow qgG_n$, and (c) $gg\rightarrow ggG_n$ which
contribute to the dijet plus graviton production process,
$pp\rightarrow jjG_n+{\rm anything}$. The gravitons are emitted
from each of the solid points in the diagrams.}}

The signal and background are simulated at the parton level 
with full tree level matrix elements. The amplitudes are 
calculated by the helicity amplitude technique~\cite{helamp},
and we have added all the relevant HELAS subroutines~\cite{helas} 
for the massive graviton and its interactions based on the 
effective Lagrangian of Eq.~(\ref{IL}). For the background, we 
used the codes  based on Ref.~\cite{vjets} and checked by 
MadGraph/MadEvent~\cite{madgraph}. For the signal, we have 
performed two independent calculations to check each other.
Firstly, we wrote our simulation codes based on the ones generated
for calculations of graviton radiation at linear
colliders~\cite{Konar:2005bd}, which matches closely
with similar other calculations~\cite{Han:1999ne}. Secondly, 
we have also implemented ADD spin-2 gravitons into MadGraph/MadEvent.
We find agreement between the two independent calculations.

Additional checks were carried out for the Born level amplitude. The
most useful check is provided by the Ward identities arising from
general coordinate invariance, which constitutes an essential
feature of any theory involving gravity. We can write the amplitude
for the emission of any graviton in the form
\begin{equation}
A_n(k,p_i) = T^{\mu\nu}(k,p_i) ~\epsilon^{(\vec{n})*}_{\mu\nu}(k) \ ,
\end{equation}
where $p_i$ are the momenta of the external SM particles,
$\epsilon^{(\vec{n})}_{\mu\nu}(k)$ is the polarization tensor for
the $\vec{n}$-th (massive) graviton mode with its momentum $k$. The tensor
$T^{\mu\nu}(k,p_i)$ is the same for all the graviton modes,
including the massless mode $\epsilon^{(0)}_{\mu\nu}(k)$, which is
the graviton of the four-dimensional Einstein gravity. This must now
satisfy the Ward identities
\begin{equation}
k^\mu T_{\mu\nu}(k,p_i)
= k^\nu T_{\mu\nu}(k,p_i) = 0 \ ,
\end{equation}
where we note that
\begin{eqnarray}
T^{\mu\nu}(k,p_i) = \sum_{j = 1}^{N} T_j^{\mu\nu}(k,p_i)
\end{eqnarray}
with $j$ indicating the $j$-th diagram, as above.  The consistency
check therefore requires a perfect cancellation between Feynman 
graphs, for each choice of
$\mu$ or $\nu$, which is highly sensitive to errors in signs and
factors. Our numerical check in Ward identities confirms the
cancellation up to the expected accuracy of our numerical programs.

Once dealing with this effective low-energy theory, one concern is
its behavior above the ADD fundamental scale ($M_s$). One unitarity 
criterion which we have implemented is that the tower of gravitons 
being produced does not extend in mass beyond the ADD scale ($M_{G_n} <
M_s$). Moreover, we will also present the results with a hard
truncation scheme, by setting the cut $Q_{{\rm truncation}}<M_s$,
where the truncation parameter is set as the root of the partonic
center-of-mass energy,
\begin{eqnarray}
Q_{{\rm truncation}} = \sqrt{\hat{s}},
\end{eqnarray}
which is a quite conservative choice.

%%%%%%%%%%%%%%%%%%%%%%%%%%%%%%%%%%%%%%%%%%%%%%%%%%%%%%%%%%%%%%%%%%%%%%%%%%%%%%%%%

\section{Results and Discussions}
In the tree level numerical calculations, we identify massless partons
with jets which must satisfy the angular cuts 
\begin{eqnarray}
&& \Delta R_{jj}= \sqrt{\Delta \eta^2+\Delta \phi^2}> 0.7 \,,\, \qquad
|\eta_{j}|< 4.5\; .\label{cut1}
\end{eqnarray}
Here $\eta$ is the pseudorapidity of the jets and $\phi$ is the
azimuthal angle around the beam direction. Unless specified otherwise we
further require
\begin{eqnarray}
&& P_T^{j}>6\;{\rm GeV}\times\sqrt{P_T^{{\rm miss}}/1\;{\rm GeV}},\label{ptmin}\\
&& P_T^{\rm miss}> 1~{\rm TeV}\; .\label{ptmiscut}
\end{eqnarray}
We employ CTEQ6L1 parton distribution functions
(PDF)~\cite{Pumplin:2002vw} throughout, with the factorization scale
chosen as $\mu_f = {\rm min}(P_T )$ of the jets which satisfy the
above cuts. The QCD coupling is set to the geometric mean
value, $\alpha_s=\sqrt{\alpha_s(P_T^{j_1}) \; \alpha_s(P_T^{j_2}})$.
For the ADD parameter, we first focus on the  $\delta=4$ and
$M_s=5$~TeV case in Figs.~\ref{fig:2}-\ref{fig:4},
and then discuss the ADD scale sensitivity and present the differential
distributions for $\delta=3,4,5,6$ cases, respectively. For
$\delta>6$ large extra dimensions, the total cross sections for
real graviton production are smaller and not easy to detect at the
LHC, thus we will not discuss them here.

Notice that throughout our calculations, we follow the
notation of Ref.~\cite{Feynr1} which differs from the one in
Ref.~\cite{Feynr2} mainly by a different factor in the relation 
between $R$ and $M_s$ in (4+$\delta$)-dimensional space.
Though this factor is crucial in comparing results and quantifying
discovery potentials, one can simply convert results from one
notation to the other by multiplying a $\delta$ - dependent
factor.

%%%%%%%%%%%%%% Begin Figure  %%%%%%%%%%%%%%%%%%%%%%%%%%%%%%%%%%%%%%%%%%
\FIGURE{
\includegraphics[width=0.49\textwidth]{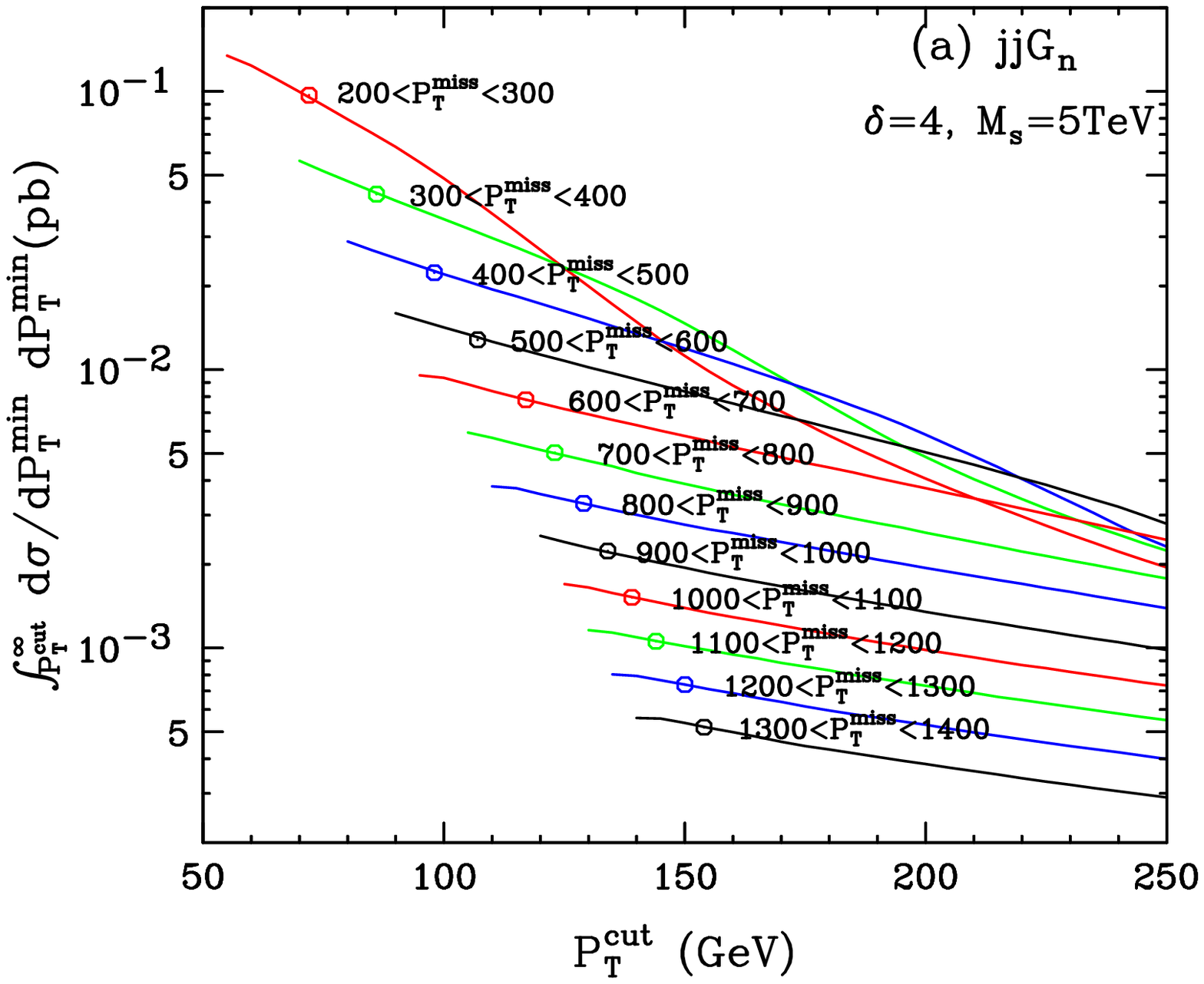}
\includegraphics[width=0.49\textwidth]{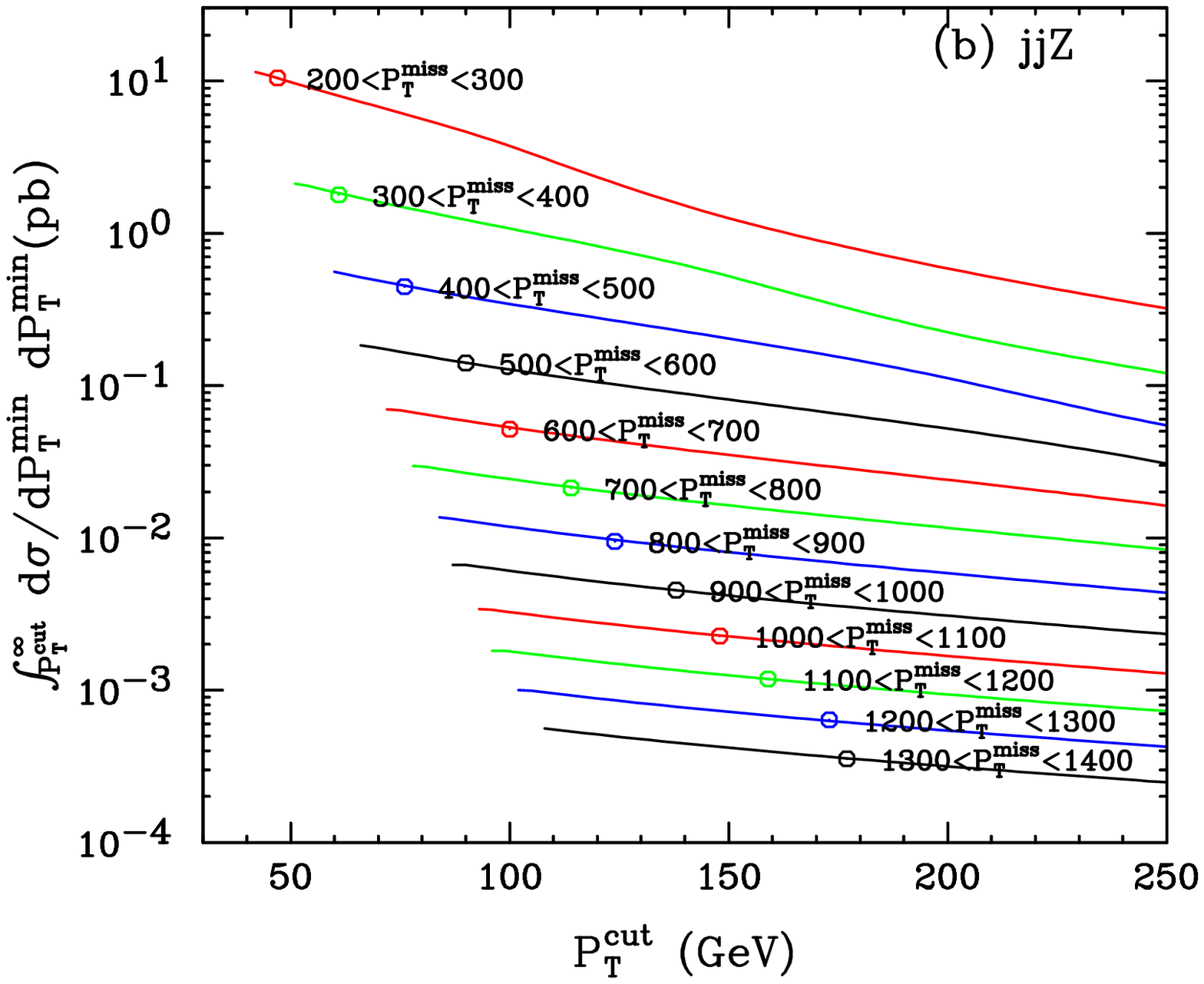}
\caption{\label{fig:2} $P_T^j$ cut dependence of the dijet cross
sections for signal (a) and background (b) at the LHC in various
missing transverse momentum bins when applying the cuts of
Eq.~(\ref{cut1}). The open circles show the monojet cross section in
 the same missing $P_T$ bin.  }}

\FIGURE{
\includegraphics[width=0.49\textwidth]{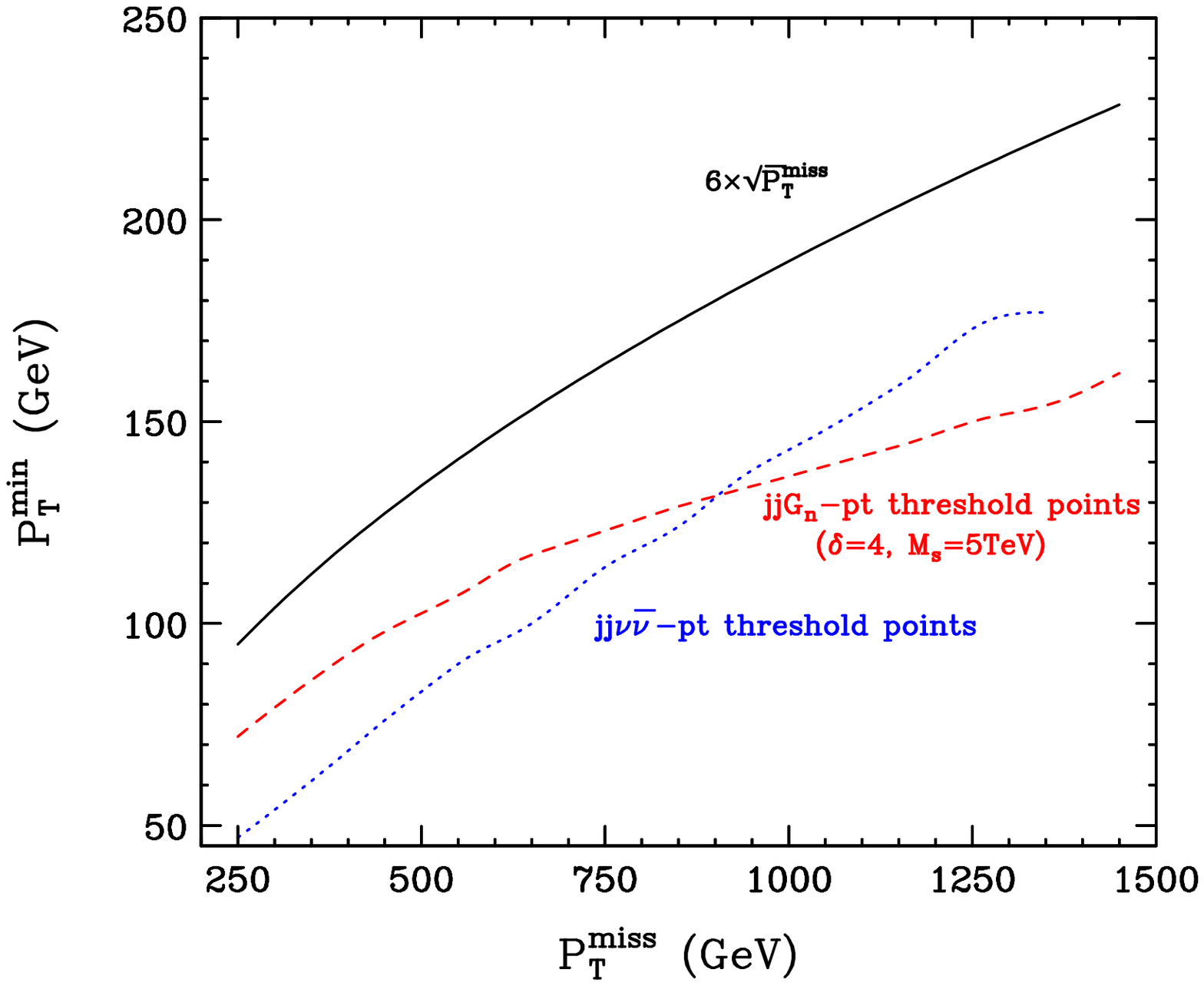}
\caption{\label{fig:3} Missing transverse momentum dependence of the 
$P_T^{{\rm cut}}$ value of equal 2-jet and 1-jet cross sections.
% The dependence of $P_T$ threshold on the missing transverse momentum. 
Our jet selection cut Eq.~(\ref{ptmin}) is also presented.}}

The  $P_T^{j}$ cut of Eq.~(\ref{ptmin}) is chosen to insure perturbative
ordering of the tree level cross sections throughout phase space,
i.e. we want to keep the dijet cross sections below the corresponding 
1-jet inclusive cross sections. In order to motivate 
our choice, we show, in Fig.~\ref{fig:2}, the $P_T^j$ cut
dependence of the total cross sections for the dijet plus missing
$P_T$ events at the LHC. Fig.~2(a) is for the signal process
$pp\rightarrow jjG_nX$, and Fig.~2(b) is for the background process
$pp\rightarrow jj(Z\rightarrow\nu\bar{\nu})X$. Each line shows the
dijet cross section for the missing $P_T$ in a 100~GeV bin between
200~GeV and 1400~GeV. The open circle along the lines shows the
corresponding monojet cross section in the same missing $P_T$ bin.
In smaller $P^{{\rm miss}}_T$ bins, the $jjG_n$ total cross sections
drop faster, with increasing $P_T^{j}$ cut, than the background ones, which is due to 
the soft and 
collinear $Z$ boson emission along one of the QCD jet directions in the $Zjj$ background.

The $P_T^{{\rm min}}$ values at open circle points in Figs.~2(a) and
(b) tell the jet $P_T$ threshold below which the dijet cross section
is larger than the monojet cross section, and hence our perturbative
results cannot be trusted. Below these threshold values one should
expect multiple soft jet emission to appear. For a missing $P_T$ of 1~TeV,
for example, gluons with $P_T\lsim 140$~GeV are in the soft range, and 
several such ``soft'' gluon jets are expected in a typical graviton or 
background event. Since these gluons are readily observable as distinct 
jets in the experiment, an actual monojet event with missing 
transverse momentum in the TeV range and no additional jets 
with $p_T\gsim 30$~GeV, is a very rare event.

%%%%%%%%%%%%%%%%%%%%%%%%%%%%%%%%%%%%%%%%%%%%%%%%%%%%%%%%%%%%%%%%%%%
\FIGURE{
\includegraphics[width=0.49\textwidth]{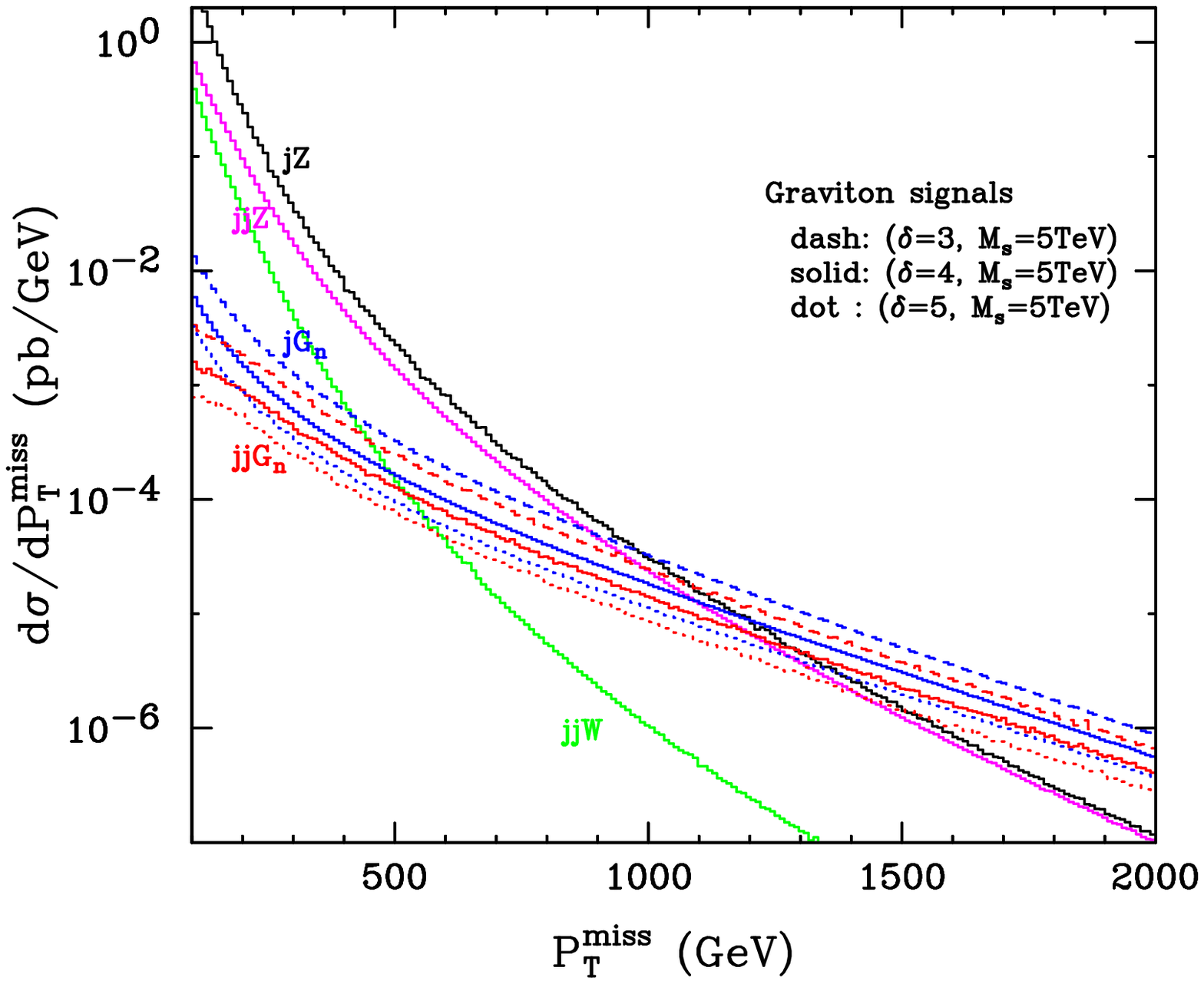}
\caption{\label{fig:4} $P^{{\rm miss}}_T$ dependence of the total
cross sections for the signal and background when applying the cuts
of Eqs. (\ref{cut1}), (\ref{ptmin}).}}

To get results which are perturbatively
reliable, we introduce the missing $P_T$ dependent jet selection cut
of Eq.~(\ref{ptmin}) such that the dijet to monojet cross section
ratio is always smaller than unity, while keeping as many dijet
events as possible. We show in Fig.~3 our jet $P_T$ cut of
Eq.~(\ref{ptmin}), together with the $P_T$ threshold values for the
signal (dashed red) and the background (dotted blue) above which the 
dijet cross
section is smaller than the monojet one. We find from this figure
that a larger fraction of graviton dijet events survives the cut for 
$P^{{\rm miss}}_T \lesssim 900$~GeV, while $Z\,\,+$~jet(s) events obtain
higher jet multiplicities at higher $P^{{\rm miss}}_T$. This is partly
because of the higher hard scattering scale of the graviton events
at small $P^{{\rm miss}}_T$, and partly because of the importance of
collinear $Z$ boson emission in high $P^{{\rm miss}}_T$ background
events.

%%%%%%%%%%%%%%%%%%%%%%%%%%%%%%%%%%%%%%%%%%%%%%%%%%%%%%%%%%%%%%%%%%%%
\FIGURE{
\includegraphics[width=0.49\textwidth]{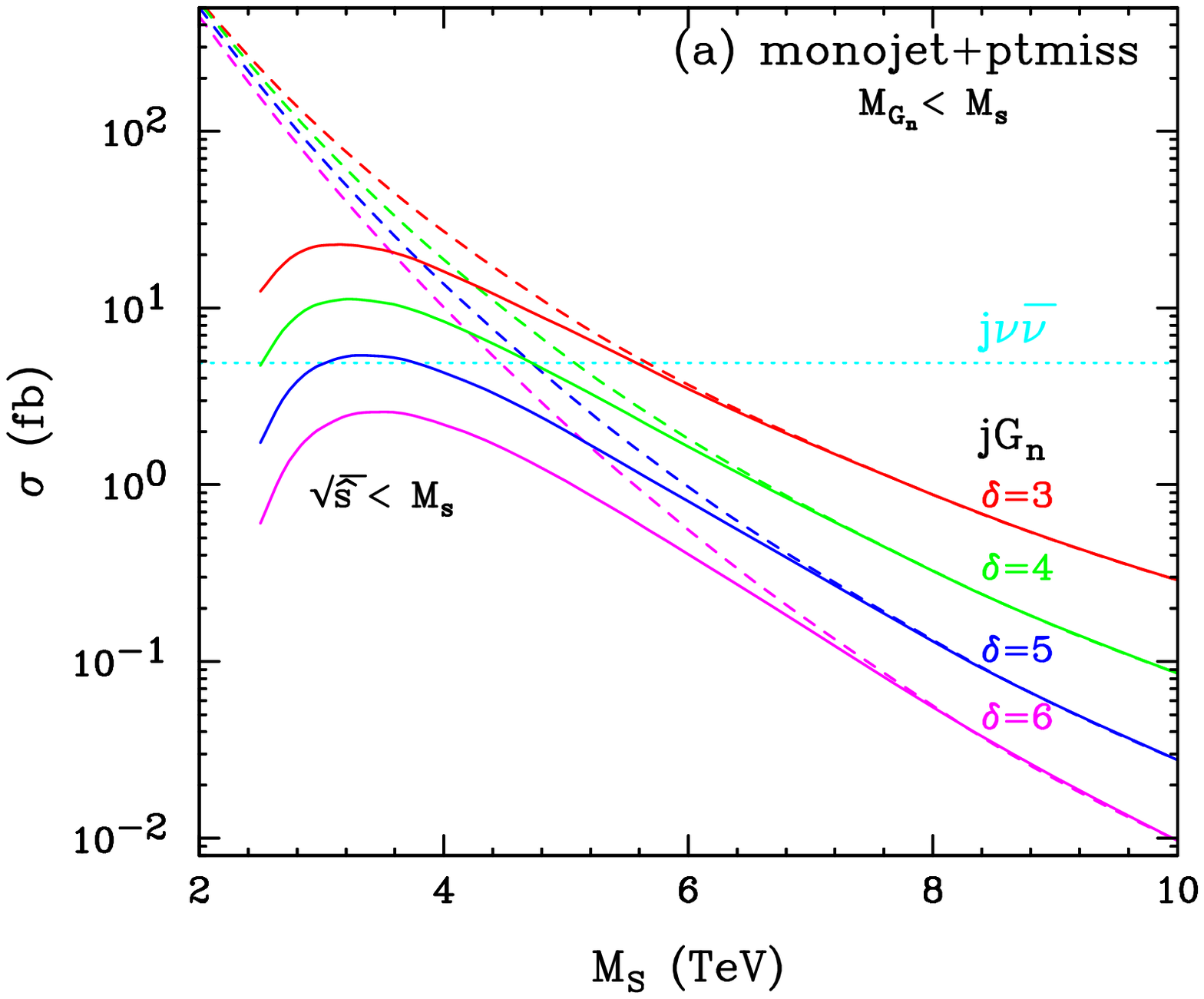}
\includegraphics[width=0.49\textwidth]{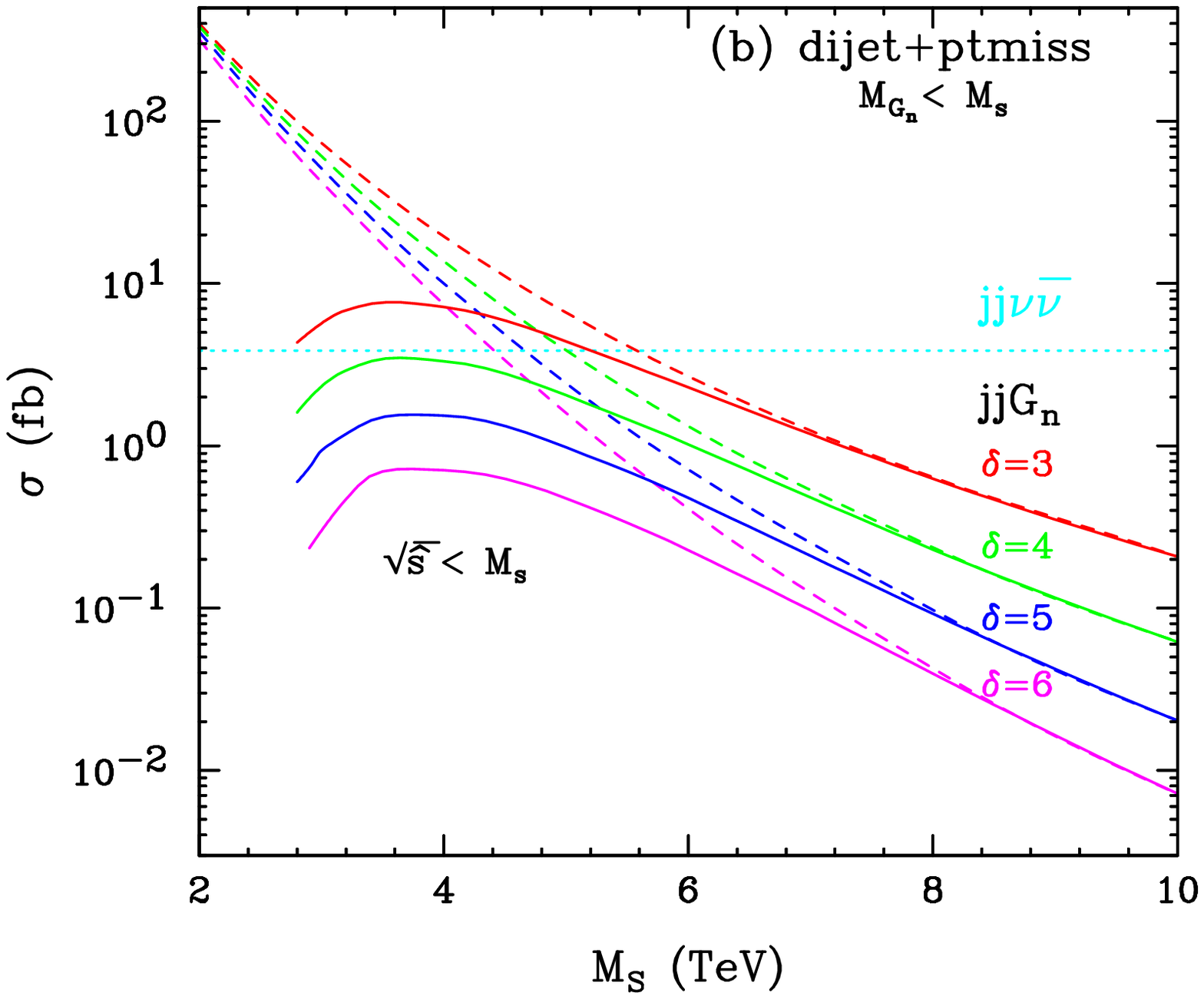}
\caption{\label{fig:5} The dependence on ADD scale $M_s$ of the total
cross sections for $jG_n$ (a) and $jjG_n$ (b) productions at the
LHC. The SM background results also have been plotted. The dashed
lines are for $M_{G_n}< M_s$, while the solid lines are for
$\sqrt{\hat{s}}<M_s$. Event selection criteria chosen as in
Eqs.~(\ref{cut1}) - (\ref{ptmiscut}).  }}

In Fig.~\ref{fig:4}, we show the $P^{{\rm miss}}_T$ spectrum for
both the 1-jet and 2-jets signal and background processes. One finds 
that the $P^{{\rm miss}}_T> 1~{\rm TeV}$
requirement in Eq.~(\ref{ptmiscut}) reduces the background
sufficiently. Also included is the contribution coming from the
$Wjj$ background which falls sharply for large $P^{{\rm miss}}_T$ and
which yields an additional background contribution of 
0.15~fb above $P^{{\rm miss}}_T = 1~{\rm TeV}$, i.e. it is negligible
compared the the $Zjj$ background.
A further improvement of the signal to background
ratio is possible by tightening the $P^{{\rm miss}}_T$  cut, but
this will not be pursued in the following.

%%%%%%%%%%%%%%%%%%%%%%%%%%%%%%%%%%%%%%%%%%%%%%%%%%%%%%%%%%%%%%%%%%%%%%%%%%
\FIGURE{
\includegraphics[width=0.49\textwidth]{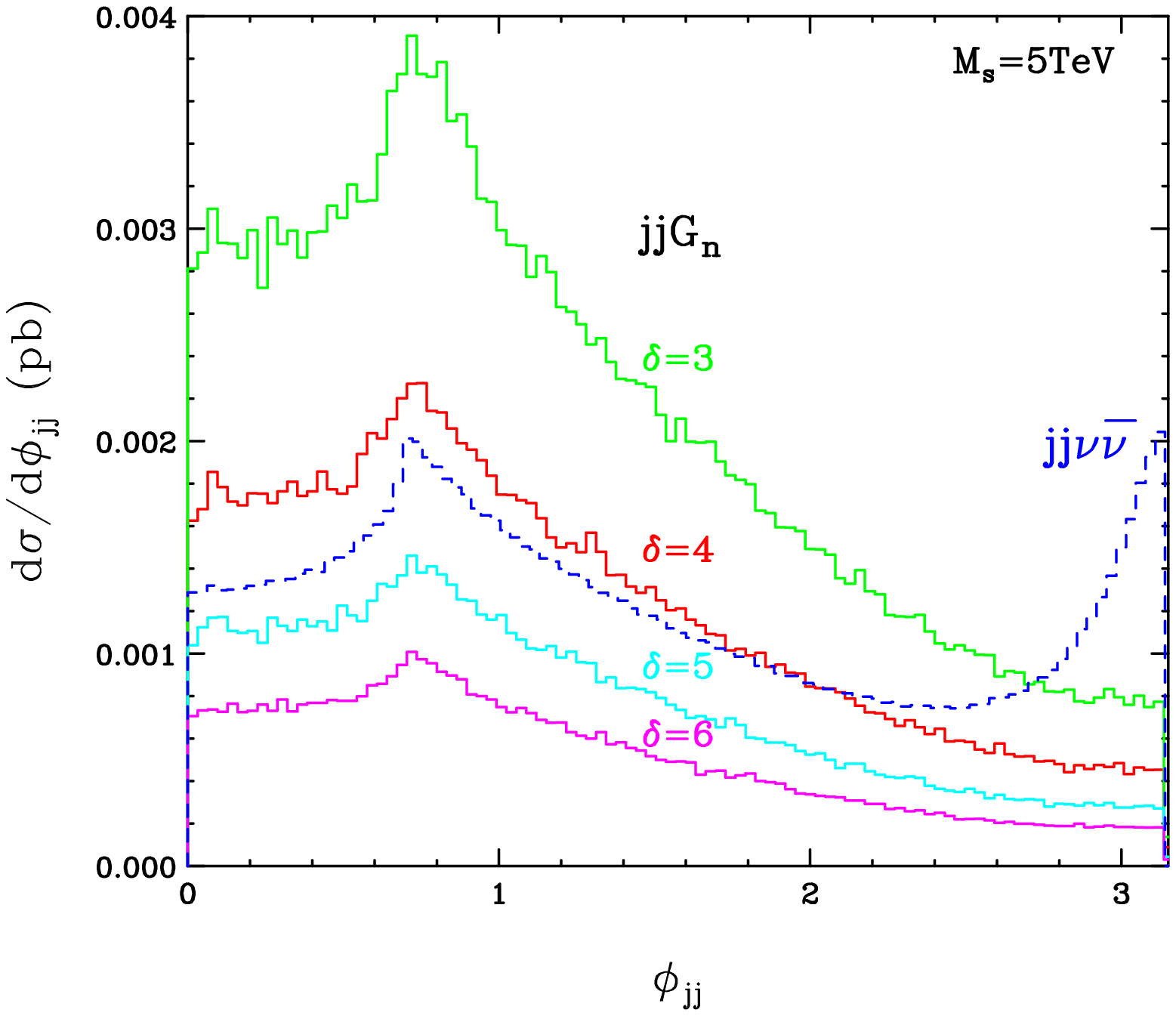}
\includegraphics[width=0.49\textwidth]{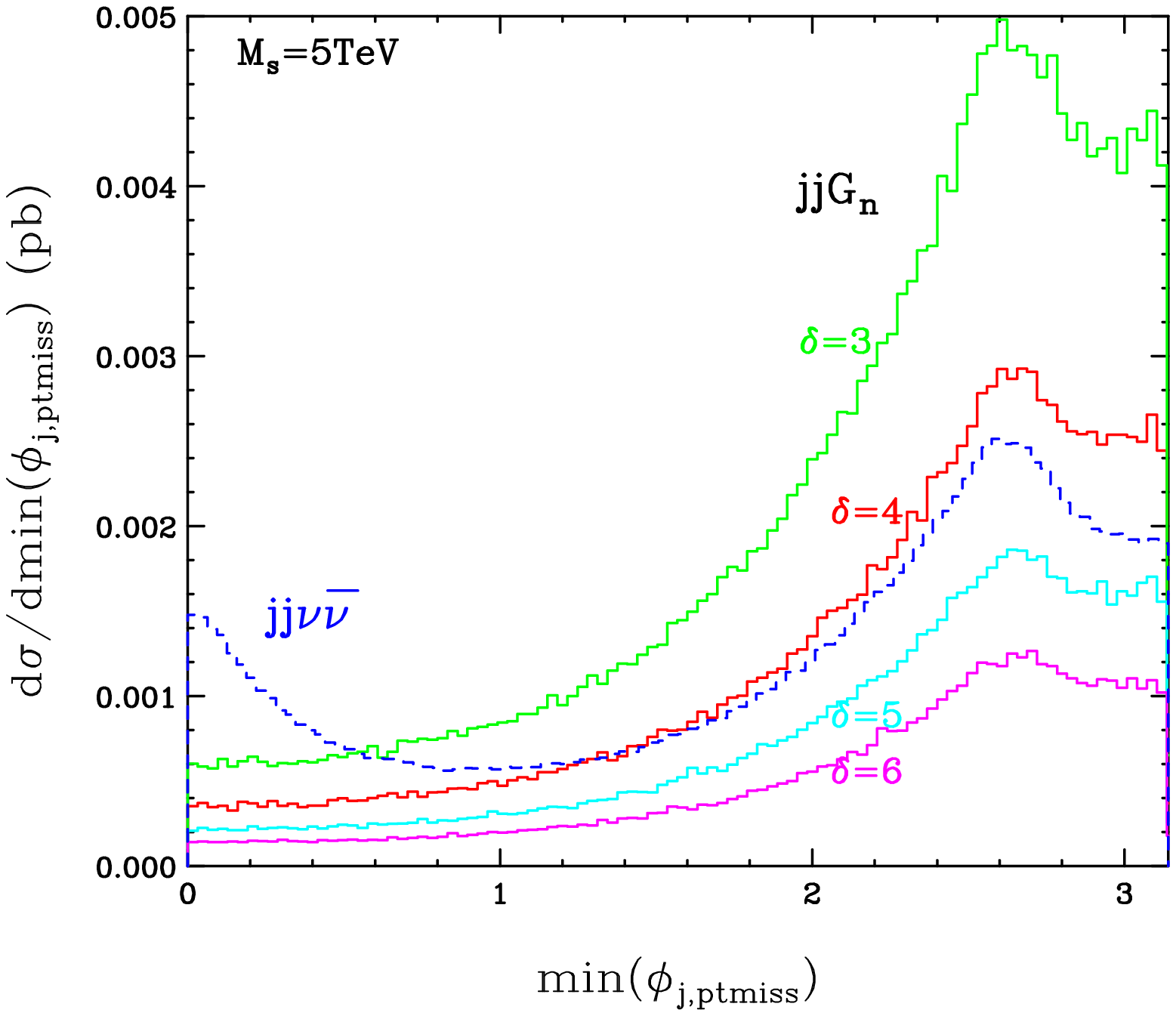}
\caption{\label{fig:6} Distributions of the azimuthal angle
 separation between the two jets (left), and the minimal azimuthal
 angle separation between the two jets and missing transverse momentum
 (right), for background and graviton signal.  }}

%%%%%%%%%%%%%%%%%%%%%%%%%%%%%%%%%%%%%%%%%%%%%%%%%%%%%%%%%%%%%%%%%%%%%
\TABLE{
\centering
\begin{tabular}{ccccc}
\hline\hline
   & Max $M_s$  sensitivity  & Max $M_s$  sensitivity\\
   & ${\cal L}=$100~${\rm fb}^{-1}$ & ${\cal L}=$100~${\rm fb}^{-1}$ \\
$\delta$   & No truncation & Hard truncation  \\
\hline
3 & 6.4 (6.6) TeV & 6.3 (6.5)  TeV \\
4 & 5.6 (5.7)~~~~~~  &  5.1 (5.5)~~~~~~ \\
5 & 5.2 (5.3)~~~~~~  &  - \,\,\,\,(4.8)~~~~~~ \\
6 & 4.9 (5.0)~~~~~~  &  - \,\,\,\,(3.6)~~~~~~ \\
\hline\hline
\end{tabular}
\caption{\label{tab} Maximum ADD scale $M_s$ sensitivity which can be reached
by studying the 2-jet (1-jet) and missing transverse momentum signal
at the LHC, with integrated luminosity ${\cal L}=$100~fb$^{-1}$,
assuming the systematic error to be $10\%$.}}

In Fig.~\ref{fig:5}, we present the ADD scale dependence of the
total cross sections for $jG_n$ and $jjG_n$ production, for the 
$\delta=3,4,5,6$ cases. The SM $j\nu\bar{\nu}$ and 
$jj\nu\bar{\nu}$ background results are also plotted. Our results for $jG_n$
production with the hard truncation scheme agree with the results in
Ref.~\cite{Feynr1} within about 5 percent, which may
be due to different PDF and scale choices. However, the results
without truncation have larger differences especially at small
$M_s$, because the unitarity criterion $M_{G_n} < M_s$ is
used in our paper as mentioned above, while not in
Ref.~\cite{Feynr1}. We have also performed the same sensitivity
analysis as in Ref.~\cite{Feynr1}, considering the integrated
luminosity ${\cal L}=100$~fb$^{-1}$, where the systematic error in the
background (assumed to be $10\%$) dominates over the statistical
error.  The sensitivity range is defined by
\begin{eqnarray}
\sigma_{jjG_n}
(\sigma_{jG_n})>5\times10\%\times\sigma_{{\rm background}}=1.93\,\,
(2.45)\, {\rm fb}.
\end{eqnarray}
The resulting max $M_s$ sensitivity results are shown in table.~I.
The 2-jet sensitivity is only slightly lower than for the
1-jet case. Moreover, the larger $\delta$ is, the sooner the
non-perturbative region is reached, thus the larger is the
difference between max $M_s$ sensitivities in no truncation and hard
truncation cases.

The most distinct difference between the signal and background in
the dijet plus missing $P_T$ events is found in azimuthal
angle correlations between the 2 jets and the missing transverse
momentum. The $\phi_{jj}$ and ${\rm min}(\phi_{j,P_T^{\rm miss}})$
distributions with the cuts (\ref{cut1})-(\ref{ptmiscut}) are shown 
in Figs.~\ref{fig:6}, for $M_s=5$~TeV and $\delta=3,4,5,6$,
respectively. The $Zjj$ background shows a clear enhancement for back to back jets, 
reflecting collinear $Z$ emission along the direction of one of the
jets. Due to the heavier masses of the typical graviton KK modes, such
collinear ``jet fragmentation'' contributions are absent for the signal.
This significant difference of the azimuthal angle distributions can
provide a powerful tool to test for heavy graviton emission: the relative
suppression of the jet fragmentation contribution in the data would be a
direct sign for a very massive object as a source of the missing
transverse momentum. 

%%%%%%%%%%%%%%%%%%%%%%%%%%%%%%%%%%%%%%%%%%%%%%%%%%%%%%%%%%%%%%%%%%%%%%%%%%
\FIGURE{
\includegraphics[width=0.49\textwidth]{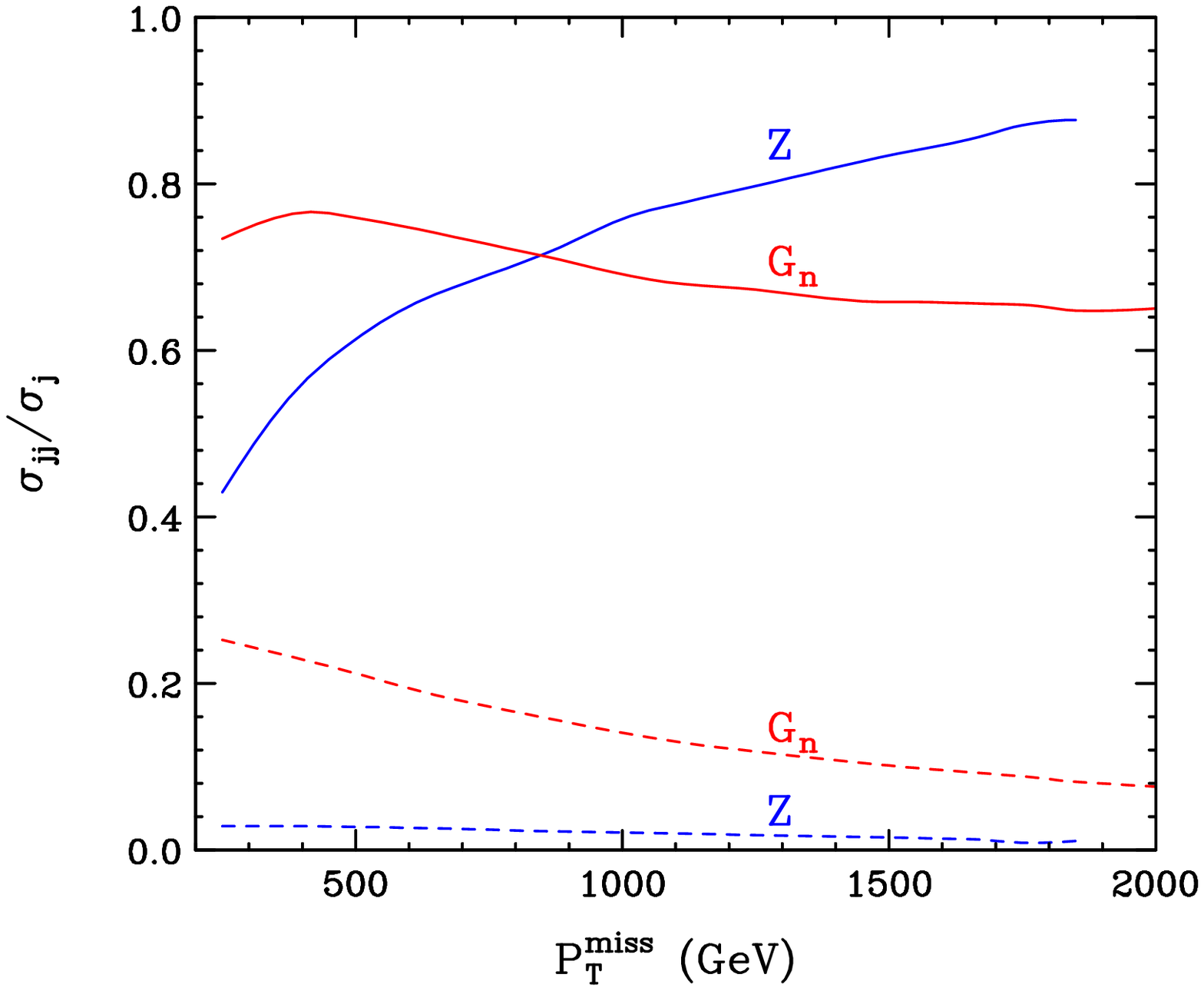}
\caption{\label{fig:7} The 2-jet over 1-jet ratio for signal and
 background as a function of $P^{{\rm miss}}_T$, with or without the
 cut $|\Delta \eta_{jj}|>$ 2 and $|\phi _{jj}-\pi|>$ 0.7. Additional
 cuts of Eqs. (\ref{cut1}), (\ref{ptmin}) applied.}}

Finally, we would like to comment on the ratio of the 2-jet
over the 1-jet rate. Since the typical graviton KK modes are much
heavier than the $Z$ boson, thus providing for a much harder event, one
would expect naively that the 
ratio for the signal should always be larger than the one for the
background. However, that is not the case as can be seen from the
results in Fig.~\ref{fig:4}, especially at large $P^{{\rm miss}}_T$.
The reason can be traced back to final state parton emission, which is
more prominent for the SM background. Based on this observation, we
show, in Fig.~\ref{fig:7}, the 2-jet over 1-jet ratio for signal and
background as a function of $P^{{\rm miss}}_T$. The upper two curves
are from the results in Fig.~\ref{fig:4}. For the lower two
curves, we set additional cuts $|\Delta \eta_{jj}|>$ 2 and $|\phi
_{jj}-\pi|>$ 0.7, in order to reduce final state parton emissions. One
now finds a higher dijet fraction for the graviton events, largely
from initial state radiation, reflecting the harder collision scale of
the graviton emission events at all missing $P_T$.

%%%%%%%%%%%%%%%%%%%%%%%%%%%%%%%%%%%%%%%%%%%%%%%%%%%%%%%%%%%%%%%%%%%%%%%%%%%%%

\section{Summary}

The production of stable Kaluza-Klein gravitons at the LHC can 
be observed when they are produced at large transverse momentum, giving 
rise to a large missing $P_T$ signature. For $P_T^{\rm miss}$ of order 
1~TeV or larger, the signature will rarely be a monojet signal, however, 
because multiple ``soft'' gluon emission will produce events with 
several jets balancing the transverse momentum of the graviton. We 
have calculated the order $\alpha_s^2$ graviton plus dijet, $jjG_n$, cross 
section at the LHC and find that it saturates the leading order monojet
cross section for additional ``soft'' jet $P_T$ in the 100 to 150 GeV 
range, thus establishing the typical scale for multiple jet emission.

The $P_T^{\rm miss}$ distribution of the $jjG_n$ cross section is strongly 
influenced by the minimal jet $p_T$ requirements imposed on the partons.
We note that defining the dijet cross section with a typical constant jet 
$P_T$ cut, independent of the hardness of the event, will invariably lead to
the cross section not being trustworthy at sufficiently high 
$P_T^{\rm miss}$, where saturation of the LO monojet cross section happens
at higher jet $P_T$ already. Not considering this effect, via a sliding
jet $P_T$ cut as in Eq.~(\ref{ptmin}), may produce an unrealistically hard
missing transverse momentum distribution, which can lead to an overestimate 
of the LHC sensitivity to graviton production.

In addition to the order $\alpha_s^2$ process, $jjG_n$ production via gluon
exchange, we have also calculated the corresponding electroweak process 
$qq\to qqG_n$ via weak boson fusion. However, the cross section for the latter
is always strongly suppressed. Even with typical weak boson fusion cuts we 
have been unable to find phase space regions where the electroweak process
contributes to overall $jjG_n$ production at more than the percent level,
while also being visible above the SM background. We conclude that 
weak boson fusion
is not a promising process for Kaluza-Klein graviton production at the LHC.

The multijet characteristics described above for the signal are also expected
for the dominant SM background, $Zjj$ production with subsequent decay of the 
$Z$ boson to neutrino pairs. The multijet features are simply a reflection 
of the hardness of the event, as specified by the large missing 
transverse momentum. We have found one feature, however, which distinguishes 
signal and background processes. For missing $P_T$ in the TeV range the $Z$
mass becomes negligible and jet fragmentation into a collinear $Z$ becomes 
an important part of the SM background. This contribution is most 
readily seen in the azimuthal angle correlations of the jets, with a sizable
fraction of nearly back-to-back dijet events. The large average mass of 
the produced gravitons strongly suppresses such a contribution for the signal.
The resulting distinctly different azimuthal angle distributions may be useful
to verify that an observed excess of high $P_T^{\rm miss}$ events is 
indeed due to the production of an invisible very massive particle.

%%%%%%%%%%%%%%%%%%%%%%%%%%%%%%%%%%%%%%%%%%%%%%%%%%%%%%%%%%%%%%%%%%%%%%%%

\acknowledgments 

K. Hagiwara wishes to thank Michael Peskin for
interesting discussions. P. Konar thanks Biswarup Mukhopadhyaya for 
his helpful suggestions. Q. Li and K. Mawatari would like to thank the
KEK theory group for its warm hospitality. This work is supported in
part by the Deutsche Forschungsgemeinschaft under SFB/TR-9
``Computergest\"utzte Theoretische Teilchenphysik'', the core
university program of JSPS, and the Grant-in-Aid for Scientific
Research (No. 17540281) of MEXT, Japan and US Department of Energy 
under grant DE-FG02-97ER41029.

The Feynman diagrams in this paper were drawn using
Jaxodraw~\cite{jaxo}.

%%%%%%%%%%%%%% Begin References %%%%%%%%%%%%%%%%%%%%%%%%%%%%%%%%%%%%%%%%

\end{document}